\documentclass[pra,print,superscriptaddress,twocolumn]{revtex4}
\usepackage{amssymb}
\usepackage{amsmath}
\usepackage{epsfig}
\usepackage{color}
\usepackage{graphics, graphicx}
\usepackage{bbold}
\usepackage{psfrag}
\usepackage{mathcomp}
\usepackage{subfigure}
\usepackage{verbatim}
\usepackage{color}
\usepackage[colorlinks,citecolor=blue]{hyperref}

\begin{document}

\title{Superfluid-Mott-insulator phase transition of light in a two-mode
cavity array with ultrastrong coupling}
\author{Jingtao Fan}
\affiliation{State Key Laboratory of Quantum Optics and Quantum Optics Devices, Institute
of Laser Spectroscopy, Shanxi University, Taiyuan 030006, China}
\affiliation{Collaborative Innovation Center of Extreme Optics, Shanxi University,
Taiyuan 030006, China}
\author{Yuanwei Zhang}
\affiliation{College of Physics and Electronic Engineering, Sichuan Normal University,
Chengdu 610068, China}
\author{Lirong Wang}
\affiliation{State Key Laboratory of Quantum Optics and Quantum Optics Devices, Institute
of Laser Spectroscopy, Shanxi University, Taiyuan 030006, China}
\affiliation{Collaborative Innovation Center of Extreme Optics, Shanxi University,
Taiyuan 030006, China}
\author{Feng Mei}
\thanks{raulmei@163.com}
\affiliation{State Key Laboratory of Quantum Optics and Quantum Optics Devices, Institute
of Laser Spectroscopy, Shanxi University, Taiyuan 030006, China}
\affiliation{Collaborative Innovation Center of Extreme Optics, Shanxi University,
Taiyuan 030006, China}
\author{Gang Chen}
\thanks{chengang971@163.com}
\affiliation{State Key Laboratory of Quantum Optics and Quantum Optics Devices, Institute
of Laser Spectroscopy, Shanxi University, Taiyuan 030006, China}
\affiliation{Collaborative Innovation Center of Extreme Optics, Shanxi University,
Taiyuan 030006, China}
\author{Suotang Jia}
\affiliation{State Key Laboratory of Quantum Optics and Quantum Optics Devices, Institute
of Laser Spectroscopy, Shanxi University, Taiyuan 030006, China}
\affiliation{Collaborative Innovation Center of Extreme Optics, Shanxi University,
Taiyuan 030006, China}

\begin{abstract}
In this paper we construct a new type of cavity array, in each cavity of
which multiple two-level atoms interact with two independent photon modes.
This system can be totally governed by a two-mode Dicke-lattice model, which
includes all of the counter-rotating terms and therefore works well in the
ultrastrong coupling regime achieved in recent experiments. Attributed to
its special atom-photon coupling scheme, this model supports a global
conserved excitation and a continuous $U(1)$ symmetry, rather than the
discrete $Z_{2}$ symmetry in the standard Dicke-lattice model. This distinct
change of symmetry via adding an extra photon mode strongly impacts the
nature of photon localization/delocalization behavior. Specifically, the
atom-photon interaction features stable Mott-lobe structures of photons and
a second-order superfluid-Mott-insulator phase transition, which share
similarities with the Jaynes-Cummings-lattice and Bose-Hubbard models. More
interestingly, the Mott-lobe structures predicted here depend crucially on
the atom number of each site. We also show that our model can be mapped into
a continuous $XX$ spin model. Finally, we propose a scheme to implement the
introduced cavity array in circuit quantum electrodynamics. This work
broadens our understanding of strongly-correlated photons.\newline
\end{abstract}

\maketitle

\section{Introduction}

Photons are excellent information carriers in nature, and generally pass
through each other without consequence. The realization of coherent
manipulation and controlling of photons allows us to achieve photon quantum
information processing \cite{TEN14} as well as to explore exotic many-body
phenomena of photons \cite{IC13}. Cavity array \cite%
{MJHF08,AAH12,SS13,IMG14,CN17}, in which each single-mode cavity interacts
with a two-level atom, is a promising platform to accomplish the required
target and has now been considered extensively \cite%
{ADG06,MFA06,MJH07,DGA07,MIM08,SS09,JK09,KT13,GK13,KKM13,SF14,BB14,MB15,ZY15,KS15,MS16,HMJ16,AM16,ALCH16}%
. On one hand, this platform has a novel interplay between strong local
nonlinearities and photon hopping of the nearest-neighbor cavities, which
has a phenomenological analogy to those of the Bose-Hubbard model \cite%
{MPA89} realized, for example, by ultracold atoms in optical lattices \cite%
{IB08}. More importantly, compared with the condensed-matter or atomic
physics, cavity array has a unique property that the fundamental many-body
phenomena depend crucially on the intrinsic atom-photon coupling strength
\cite{MJHF08,AAH12,SS13,IMG14,CN17}.

For the weak and moderately-strong coupling regimes, the counter-rotating
terms of the single-site Hamiltonian are usually neglected by employing the
rotating-wave approximation. As a result, the property of cavity array is
governed by a Jaynes-Cummings-lattice model \cite%
{MJHF08,AAH12,SS13,IMG14,CN17}. Since this Jaynes-Cummings-lattice model
preserves a global excitation number, a series of Mott insulator (MI) phase
of photons form a lobe structure and a second-order superfluid(SF)-MI phase
transition take place across the edge of each lobe. This Mott-lobe structure
makes it a photonic counterpart of the Bose-Hubbard model \cite{MPA89},
which simulates massive bosons in lattice and also supports a similar lobe
structure. However, it should be noticed that a complete description of the
light-matter interaction should always incorporate the counter-rotating
terms, especially considering the fact that recent experiments of circuit
quantum electrodynamics (QED) have accessed the ultrastrong coupling regime
(i.e., the atom-photon coupling strength has the same order of the photon
frequency) \cite{TN10,PF10,PJ16,FY16}, in which the rotating-wave
approximation totally breaks down. In such a case, a proper description of
the system dynamics should resort to a Rabi-lattice model. Since the
counter-rotating terms in the Rabi-lattice model breaks the conservation of
excitation number, there is, in principle, no similar MI as that of the
Bose-Hubbard model and the transition between the SF and MI should be
replaced by the coherent and incoherent type \cite{MSCH12,MSCH13}. These
essential changes of equilibrium properties motivate us to ask a question:
could the Mott-lobe structure still exist even though all of the
counter-rotating terms of the atom-photon coupling are taken into consideration?

In the present paper, we try to answer this question by constructing a new
type of cavity array, in each cavity of which multiple two-level atoms
interact with two independent photon modes. This system can be totally
governed by a two-mode Dicke-lattice (TMDL) model, which includes all of the
counter-rotating terms and therefore works well in the ultrastrong coupling
regime. Unlike the Rabi-lattice model, the TMDL model has a global conserved
excitation and a continuous $U(1)$ symmetry. This distinct change of
symmetry via adding an extra photon mode induces some interesting many-body
physics of strongly-correlated photons. Specifically, the atom-photon
interaction features stable Mott-lobe structures of photons and a
second-order SF-MI phase transition, which share similarities with the
Jaynes-Cummings-lattice \cite{MJHF08,AAH12,SS13,IMG14,CN17} and Bose-Hubbard
\cite{MPA89} models. However, in contrast to these models, the Mott-lobe
structures predicted here depend crucially on the atom number of each site,
reflecting its particularity among lattice models. We also show that the
TMDL model can be mapped into a continuous $XX$ spin model under proper
parameter conditions. Finally, motivated by recent experimental achievements
of cavity array \cite{JR14,CE14,MF16} and multimode cavity \cite%
{MM11,NMS15,DCM15} in circuit QED, we propose a scheme to realize the TMDL
model in a two-mode superconducting stripline cavity array. This work
broadens our understanding of strongly-correlated photons.\newline

\section{Model and Hamiltonian}

We study a photon lattice system composed by an array of identical coupled
cavities, inside each of which multiple two-level atoms interact with two
degenerate photon modes. Such a system is governed by the TMDL Hamiltonian
\begin{equation}
\hat{H}_{\text{T}}=\sum_{j}\hat{H}_{j}^{\text{TD}}-t\sum_{\left\langle
j,k\right\rangle }\sum_{m=1,2}\hat{a}_{m,j}^{\dag }\hat{a}_{m,k},
\label{TMDM}
\end{equation}%
where the single-site Hamiltonian
\begin{eqnarray}
\hat{H}_{j}^{\text{TD}} &=&\omega \sum_{m=1,2}\hat{a}_{m,j}^{\dag }\hat{a}%
_{m,j}+\omega _{0}\hat{J}_{z,j}+  \label{TDH} \\
&&g\left[ \left( \hat{a}_{1,j}+\hat{a}_{1,j}^{\dag }\right) \hat{J}%
_{x,j}+i\left( \hat{a}_{2,j}-\hat{a}_{2,j}^{\dag }\right) \hat{J}_{y,j}%
\right] .  \notag
\end{eqnarray}%
In the Hamiltonians (\ref{TMDM}) and (\ref{TDH}), $\hat{a}_{m,j}^{\dag }$
and $\hat{a}_{m,j}$ are the creation and annihilation operators of the $m$th
photon mode of site $j$, $\hat{J}_{i,j}(i=x,y,z)=$ $\sum_{l=1}^{N}\hat{\sigma%
}_{i,j}^{l}/2$, with $\hat{\sigma}_{i,j}^{l}$ being the Pauli spin operator,
is the collective spin operator of site $j$, $\omega $ is the frequency of
the degenerate photon modes, $\omega _{0}$ is the atom resonant frequency, $%
g $ is the atom-photon coupling strength, $t$ is the hopping rate, and $%
\left\langle j,k\right\rangle $ denotes the photon hopping between the
nearest-neighbor sites $j$ and $k$.

An intriguing feature of the Hamiltonian (\ref{TDH}) is that the spin
operator couples to the two independent photon modes via its two orthogonal
components $\hat{J}_{x}$ and $\hat{J}_{y}$, respectively. Without the
coupling term $i(\hat{a}_{2,j}-\hat{a}_{2,j}^{\dag })J_{y,j}$, the
Hamiltonian (\ref{TDH}) reduces to the standard Dicke model
\begin{equation}
\hat{H}_{j}^{\text{D}}=\omega \hat{a}_{1,j}^{\dag }\hat{a}_{1,j}+\omega _{0}%
\hat{J}_{z,j}+g(\hat{a}_{1,j}+\hat{a}_{1,j}^{\dag })\hat{J}_{x,j},
\label{STAND}
\end{equation}%
and the corresponding Hamiltonian (\ref{TMDM}) is thus called the
Dicke-lattice model \cite{LJZ14} (Rabi-lattice model for $N=1$ \cite%
{HZ11,MSCH12,MSCH13,TF16,BS13}, with $N$ being the atom number of each
site). Obviously, since the rotating-wave approximation is not employed, the
TMDL model is able to completely describe potential effects arising from the
counter-rotating terms and is therefore reasonable in the ultrastrong
coupling regime, which has been achieved in current experiments of circuit
QED \cite{TN10,PF10,PJ16,FY16}.

The emergence of the so-called counter-rotating terms in the Dicke
Hamiltonian (\ref{STAND}) reduces the conservation of its excitation number,
$\hat{N}_{s,j}=$ $\hat{J}_{z,j}+\hat{a}_{1,j}^{\dag }\hat{a}_{1,j}$, to a
parity $\Pi =\exp (i\pi \hat{N}_{s,j})$. However, by introducing an extra
degenerate photon mode $\hat{a}_{2,j}$, the Hamiltonian (\ref{TDH}) exhibits
a special conserved excitation \cite{JF14}, $\hat{N}_{e,j}=\hat{J}_{z,j}+%
\hat{a}_{1,j}^{\dag }\hat{a}_{2,j}+\hat{a}_{2,j}^{\dag }\hat{a}_{1,j}$,
apart from the known conserved parity \cite{CT03}, even if the rotating-wave
approximation is not applied. When the photon hopping is triggered on, this
conserved local excitation $\hat{N}_{e,j}$ is replaced by a global one,
\begin{equation}
\hat{N}_{e}=\sum_{j}\hat{N}_{e,j}=\sum_{j}(\hat{J}_{z,j}+\hat{a}_{1,j}^{\dag
}\hat{a}_{2,j}+\hat{a}_{2,j}^{\dag }\hat{a}_{1,j}),  \label{SGE}
\end{equation}%
which manifests the $U(1)$ symmetry of the Hamiltonian (\ref{TMDM}). The
conserved global excitation $\hat{N}_{e}$ and its induced $U(1)$ symmetry
distinguish the TMDL model from the standard Dicke-lattice model (with a
discrete $Z_{2}$ symmetry and without conserved excitation). This complete
change of symmetry are expected to deeply impact the behavior of
strongly-correlated photons.\newline

\section{Ground-state phase diagram}

\begin{figure}[tp]
\includegraphics[width=8cm]{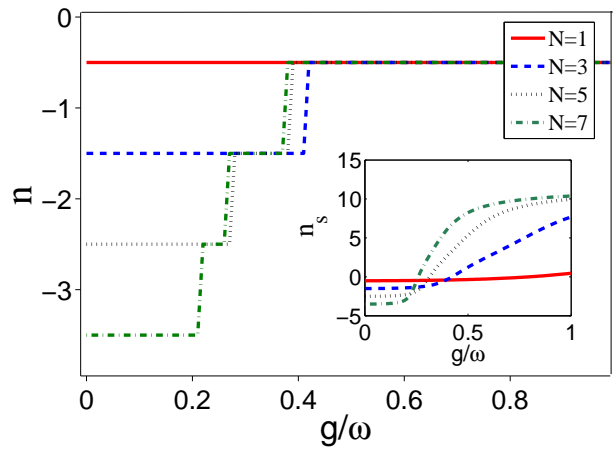}\newline
\caption{The ground-state mean excitation densities, $n=\left\langle \hat{N}%
_{e,j}\right\rangle $, of the two-mode Dicke model (\protect\ref{TDH}) as
functions of $g/\protect\omega $ for different $N$. Inset: the mean
excitation densities, $n_{s}=\left\langle \hat{N}_{s,j}\right\rangle $, of
the standard Dicke model. In these figures, we set $\protect\omega _{0}/%
\protect\omega =1$. }
\label{meanexcitationg}
\end{figure}

Since the knowledge of the single-site limit is crucial for a further
understanding of many-body physics, before proceeding, we first catch some
instructive insights into the Hamiltonian (\ref{TDH}). In the absence of the
photon hopping ($t=0$), the excitation density $\hat{N}_{e,j}$ commutes with
the Hamiltonian (\ref{TMDM}) and each eigenstate is thus characterized by a
certain excitation number. With an increasing of the system parameter, the
level-crossings of the lowest eigenstates are expected to take place,
switching a definite excitation density of the ground state. Armed with this
argument, we plot the ground-state mean excitation density, $n=\left\langle
\hat{N}_{e,j}\right\rangle $, of the single-site Hamiltonian (\ref{TDH}) as
a function of $g$ in Fig.~\ref{meanexcitationg}. The evolution of $n$
reflects a conspicuous staircase, whose jump points are associated with the
crossover points of the lowest energy levels. For $N=1$, $n$ remains a
constant, whereas when increasing $N$, the staircase appears and becomes
more and more crowded, showing that the level crossing occurs only for $%
N\geqslant 2$. This property is totally different from the standard Dicke
model (\ref{STAND}), where no staircase can be found for any $N$ [see the
insert part of Fig.~\ref{meanexcitationg}], due to the nonconservation of
its excitation density $\hat{N}_{s,j}$.

We now pay attention to the TMDL Hamiltonian (\ref{TMDM}). By applying a
mean-field decoupling approximation \cite{MPA89}, i.e., $\hat{a}_{m,j}^{\dag
}\hat{a}_{m,k}=\langle \hat{a}_{m,j}^{\dag }\rangle \hat{a}_{m,k}+\langle
\hat{a}_{m,k}\rangle \hat{a}_{m,j}^{\dag }-\langle \hat{a}_{m,j}^{\dag
}\rangle \langle \hat{a}_{m,k}\rangle $, the many-body Hamiltonian (\ref%
{TMDM}) reduces to an effective mean-field Hamiltonian%
\begin{equation}
\hat{H}_{\mathcal{MF}}=\sum_{j}\hat{H}_{j}^{\text{TD}}-zt\sum_{j,m}\left[
\psi _{m}\left( \hat{a}_{m,j}^{\dag }+\hat{a}_{m,j}\right) -\left\vert \psi
_{m}\right\vert ^{2}\right] ,  \label{HMF}
\end{equation}%
where $z$ denotes the number of nearest neighbors, and $\psi _{m}=\langle
\hat{a}_{m,j}\rangle $ ($m=1,2$) is the variational SF order parameter,
which is taken to be real for simplicity \cite{ADG06,SCR08}. $\psi _{m}$ can
be determined self-consistently by minimizing the ground-state energy $%
E(\psi _{1},\psi _{2})$ of the mean-field Hamiltonian (\ref{HMF}) \cite%
{ADG06}.

The effective mean-field Hamiltonian (\ref{HMF}) reveals an intimate
connections between the single-site Hamiltonian (\ref{TDH}) and the
many-body properties. In general, even though the global excitation $\hat{N}%
_{e}$ is a conserved quantity, the excitation number $\hat{N}_{e,j}$ of each
site does not conserve, due to the photon hopping. However, as shown in the
Hamiltonian (\ref{HMF}), if both $\psi _{1}$ and $\psi _{2}$ vanish, the
system dynamics is dominated by the single-site Hamiltonian (\ref{TDH}), and
the photons at each site are thus effectively frozen and characterized by a
specific excitation number $n$. We accordingly denote this case as a MI
phase, in which the $U(1)$ symmetry is preserved. Whereas a $U(1)$
symmetry-broken phase, associated with the breaking of the conservation of $%
\hat{N}_{e,j}$, is symbolized by a nonzero value of $\psi _{m}$ and can be
anticipated across a critical hopping rate $t_{c}(g)$. In this condition,
the photon mode $m$ governs a macroscopic coherence over the lattice and we
have a SF phase of the mode $m$. It was generally believed that the complete
inclusion of the counter-rotating terms would demolish the MI phase since
they couple states with different numbers of the dressed photons and
therefore inhibit the formation of photon blockade, which is crucially
necessary for the MI phase \cite{HZ11,MSCH12,MSCH13,TF16}. In such a case,
the notion \textquotedblleft SF/MI\textquotedblright\ should be replaced by
\textquotedblleft coherent/incoherent\textquotedblright . Nevertheless, the
TMDL model we introduced here offers a superb exception $-- $ although still
breaking the conventional conservation of $\hat{N}_{s,j}$, the
counter-rotating terms in the TMDL model preserve the hybridized two-mode
excitation $\hat{N}_{e,j}$, attributed to the special atom-photon coupling
scheme in the Hamiltonian (\ref{TDH}), and thus retain the possibility to
form the SF-MI phase transition.

\begin{figure}[tp]
\includegraphics[width=8cm]{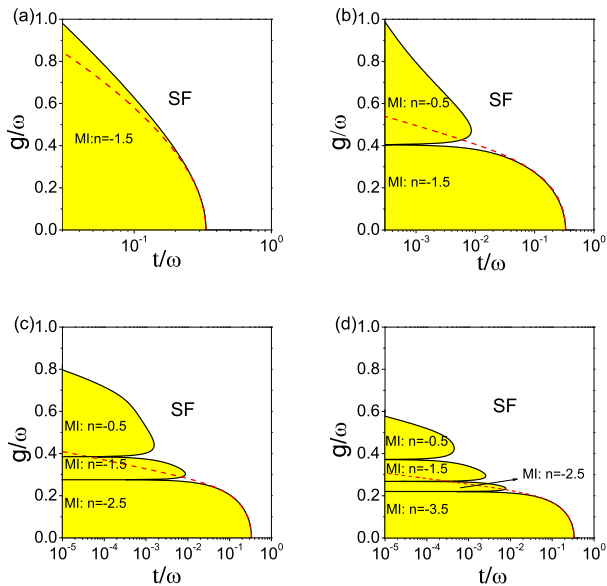}\newline
\caption{Ground-state phase diagrams of the Hamiltonian (\protect\ref{HMF})
in the $t-g$ plane, when (a) $N=1$, (b) $N=3$, (c) $N=5$, and (d) $N=7$. The
MI phase is characterized by its lobes, each of which supports a constant
mean excitation density $n=\left\langle \hat{N}_{e,j}\right\rangle $. For a
comparison, the phase boundaries of the Dicke-lattice model are also shown
by the red-dashed curves. In these figures, we set $\protect\omega _{0}/%
\protect\omega =1$.}
\label{tg}
\end{figure}

Based on above considerations, we plot the ground-state phase diagram in the
$t-g$ plane for different $N$ in Fig.~\ref{tg}. These results show two
typical phases: the $U(1)$ symmetry-preserved MI with $\psi _{1}=\psi _{2}=0$
and the symmetry-broken SF with nonzero $\psi _{1}$ and $\psi _{2}$. A
further analysis of $\psi _{m}$ near the critical point demonstrates that
the transition between these two phases is of second order. According to the
Landau's theory \cite{JPS06,KH87}, the phase boundary of such a continuous
transition can be obtained by a perturbation method, in which the
ground-state energy $E_{n}(\psi _{1},\psi _{2})$ is expanded up to second
order in $\psi _{m}$ \cite{JK09,MSCH13}. We expand $E_{n}(\psi _{1},\psi
_{2})$ of the $n$th MI phase around the critical value of the order
parameter $\psi _{m}=0$. The expanded ground-state energy in powers of $%
zt\psi $ reads
\begin{equation}
E_{n}(\psi _{1},\psi _{2})=E_{n}^{(0)}+E_{n}^{(2)}+O(tz\psi )^{4},
\label{SEO}
\end{equation}%
where the second-order energy correction
\begin{equation}
E_{n}^{(2)}=\sum_{m=1,2}(zt+z^{2}t^{2}R_{m,n})\left\vert \psi
_{m}\right\vert ^{2}+2z^{2}t^{2}T_{n}\psi _{1}\psi _{2}.  \label{ENT}
\end{equation}%
The coefficients $R_{m,n}$ and $T_{n}$ in Eq.~(\ref{ENT}) are derived from
the second-order perturbation theory by
\begin{equation}
R_{m,n}=\sum_{k\neq n}\frac{\left\vert \left\langle n\right\vert (\hat{a}%
_{m,j}+\hat{a}_{m,j}^{\dag })\left\vert k\right\rangle \right\vert ^{2}}{%
E_{n}^{(0)}-E_{k}^{(0)}},  \label{RMN}
\end{equation}%
and%
\begin{equation}
T_{n}=\sum_{k\neq n}\frac{\left[ \left\langle n\right\vert (\hat{a}_{1,j}+%
\hat{a}_{1,j}^{\dag })\left\vert k\right\rangle \left\langle k\right\vert (%
\hat{a}_{2,j}+\hat{a}_{2,j}^{\dag })\left\vert n\right\rangle +c.c\right] }{%
2(E_{n}^{(0)}-E_{k}^{(0)})}.  \label{TN}
\end{equation}%
where $E_{k}^{(0)}$ and $\left\vert k\right\rangle $ arise from the
eigenequation $\hat{H}_{j}^{\text{TD}}\left\vert k\right\rangle
=E_{k}^{(0)}\left\vert k\right\rangle $.

The critical hopping rate $t_{c}$ can be obtained by the following
procedure. (i) We first write a $2\times 2$ Hessian matrix in terms of Eq.~(%
\ref{ENT}), i.e., $\mathcal{M}_{ij}=\partial ^{2}E_{n}^{(2)}/\partial \psi
_{i}\partial \psi _{j}$, and then derive its two eigenvalues $\varepsilon
_{1}$ and $\varepsilon _{2}$. (ii) These two eigenvalues generate two
equations, $\varepsilon _{1}=0$ and $\varepsilon _{2}=0$, with respect to $t$%
. Each of these equations, say $\varepsilon _{m}=0$, supports a trivial
solution $t_{m}^{T}=0$ and a nontrivial solution $t_{m}^{N}\neq 0$. (iii)
The critical transition point is finally given by%
\begin{equation}
t_{c}=\min (t_{1}^{N},t_{2}^{N}).  \label{TC}
\end{equation}

The obtained boundaries are shown by the black solid curves in Fig.~\ref{tg}%
. The most important finding, as expected, is that the missing Mott lobes in
the standard Dicke-lattice model \cite{HZ11,MSCH12} (see the red dashed
curve in Fig.~\ref{tg}) reappear. More interestingly, our predicted
Mott-lobe structure depends crucially on the atom number $N$, which has no
counterpart in the Jaynes-Cummings-lattice \cite%
{MJHF08,AAH12,SS13,IMG14,CN17} and Bose-Hubbard \cite{MPA89} models (Note
that the $N$-dependent phase diagram for the Tavis-Cummings-lattice, which
is nothing but the Dicke-lattice after the rotating-wave approximation, has
been investigated previously \cite{SCR08,DR07,MK10}. In that case, the atom
number $N$ only slightly shifts the phase boundary of each lobe, rather than
its total structure). Specifically, when $N=1$, the atom-photon coupling
features only a single Mott lobe, as shown in Fig.~\ref{tg}(a). With the
increasing of $N$, however, more and more Mott lobes emerge, as shown in
Figs.~\ref{tg}(b)-\ref{tg}(d). This $N$-dependent behavior of the Mott lobes
is a direct legacy of the $N$-dependent staircase of $n$ governed by the
single-site Hamiltonian (\ref{TDH}). In fact, since in the MI phase, the
mean-field Hamiltonian (\ref{HMF}) equals to the single-site Hamiltonian (%
\ref{TDH}), there exists a one-to-one correspondence between Fig.~\ref%
{meanexcitationg} and Fig.~\ref{tg}. As a result, each Mott lobe is
specified by a definite mean excitation density $n$.

\begin{figure}[tp]
\includegraphics[width=8cm]{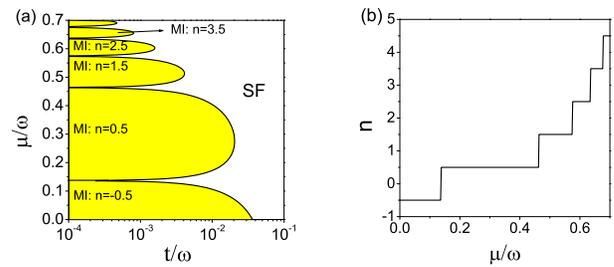}\newline
\caption{(a) Ground-state phase diagram of the Hamiltonian (\protect\ref{GCE}%
) in the $t-\protect\mu $ plane and (b) the corresponding mean excitation
density, $n=\left\langle \hat{N}_{e,j}\right\rangle $, of the single-site
limit as a function of $\protect\mu /\protect\omega $. In these figures, we
set $g/\protect\omega =\protect\omega _{0}/\protect\omega =1$ and $N=1$.}
\label{excitationmu}
\end{figure}

We emphasize that in the TMDL model, on the one hand, no chemical potential
is needed to engineer the Mott lobes, which is here stabilized by the
atom-photon coupling instead \cite{MSCH12,MSCH13}. This is in sharp contrast
to both the cases of the Jaynes-Cummings-lattice \cite%
{MJHF08,AAH12,SS13,IMG14,CN17} and Bose-Hubbard \cite{MPA89} models, which
are often studied within the framework of grand canonical ensemble where a
chemical potential is introduced to fix the (conserved) number of
excitations on the lattice \cite{ADG06,JK09}. On the other hand, the
standard Dicke- or Rabi-lattice model does not support any conserved
excitations, due to the inclusion of the counter-rotating terms. This makes
the description of grand canonical ensemble irrelevant to some extent and no
well-defined chemical potential thus exists \cite{JPS06,KH87}. However, the
conserved excitation in the TMDL model motivates us to introduce a chemical
potential $\mu $ and access a theory of grand canonical ensemble. We now
extend Eq.~(\ref{TMDM}) to the following Hamiltonian in grand canonical
ensemble:
\begin{eqnarray}
\hat{H}_{\text{G}} &=&\hat{H}_{\text{C}}-\mu \hat{N}_{e}  \label{GCE} \\
&=&\sum_{j}\hat{H}_{j}^{\text{GTD}}-t\sum_{\left\langle j,k\right\rangle
}\sum_{m=1,2}\hat{a}_{m,j}^{\dag }\hat{a}_{m,k},  \notag
\end{eqnarray}%
where the on-site two-mode Dicke Hamiltonian becomes $\hat{H}_{j}^{\text{GTD}%
}=\hat{H}_{j}^{\text{TD}}-\mu \hat{N}_{e,j}$. Following the same mean-field
theory, we plot the phase diagram in the $t-\mu $ plane in Fig.~\ref%
{excitationmu}. As shown in Fig.~\ref{excitationmu}(a), the engineered
chemical potential $\mu $ still features the Mott lobes, which is a direct
analog of those of the Bose-Hubbard model \cite{MPA89}. Once again, a clear
interpretation of this lobe structure is still based on the dynamics of the
single-site limit, which is governed by the Hamiltonian $\hat{H}_{j}^{\text{%
GTD}}$. As the chemical potential couples to a conserved quantity $\hat{N}%
_{e,j}$ in the Hamiltonian $\hat{H}_{j}^{\text{GTD}}$, the eigenstates are
independent of $\mu $, due to the simultaneous diagonalization of $\hat{H}%
_{j}^{\text{TD}}$ and $\hat{N}_{e,j}$. Thus, the ground-state competition
leads to a staircase behavior of the excitation density $\hat{N}_{e,j}$ when
varying $\mu $, as shown in Fig.~\ref{excitationmu}(b). And accordingly,
each Mott lobe in Fig.~\ref{excitationmu}(a) is characterized by the
corresponding plateaux.\newline

\section{Effective spin model: the continuous $XX$ model}

It has been well established that the Jaynes-Cummings-lattice model,
respecting a $U(1)$ symmetry, can be mapped to a continuous $XX$ spin model
(the isotropic $XY$ spin model) \cite{DGA07,JK09}, whereas the Rabi-lattice
model with the counter-rotating terms has been demonstrated to be in the
Ising universality class, owing to its discrete $Z_{2}$ symmetry \cite%
{MSCH12,MSCH13,BS13}. As revealed in this paper, however, the inclusion of
the counter-rotating terms does not always break the continuous symmetry.
Especially, for our TMDL model, the $U(1)$ symmetry associated with the
conserved excitation number is a signature of its intimate connection with
the continuous spin model. To confirm this argument, we focus on the system
dynamics in the $t-g$ plane, which is governed by the Hamiltonian (\ref{TMDM}%
). We first consider the case of $N\geqslant 2$, which supports a multi-lobe
structure in the phase diagram.

When parameters are tuned close to the degenerate point in the MI phase with
$t\ll g$, i.e., the boundary between two nearest Mott lobes, we can truncate
the Hilbert space to two of the excitation number eigenstates $\left\vert
n\right\rangle $ and $\left\vert n+1\right\rangle $, where $\left\vert
n\right\rangle $ denotes the eigenstate of the excitation density $\hat{N}%
_{e,j}$ with eigenvalue $n$ (as verified numerically below, $n$ varies only
by one across the degenerate point). Utilizing the commutation relations
between the photon annihilation operator $\hat{a}_{m,j}$ and the excitation
number $\hat{N}_{e,j}$, we can map $\hat{a}_{m,j}$ in the reduced Hilbert
space $\left\{ \left\vert n\right\rangle ,\left\vert n+1\right\rangle
\right\} $ into
\begin{equation}
\hat{a}_{1,j}\rightarrow \alpha \hat{\Sigma}_{j}^{-}+\beta \hat{\Sigma}%
_{j}^{+}\text{, \ \ \ \ }\hat{a}_{2,j}\rightarrow \alpha \hat{\Sigma}%
_{j}^{-}-\beta \hat{\Sigma}_{j}^{+},  \label{OP}
\end{equation}%
where $\hat{\Sigma}_{j}^{+}=\left\vert n\right\rangle \left\langle
n+1\right\vert $ and $\hat{\Sigma}_{j}^{-}=\left\vert n+1\right\rangle
\left\langle n\right\vert $ are the redefined Pauli spin ladder operators,
and the coefficients $\alpha $ and $\beta $ can be determined numerically
(see Appendix A for details). Therefore, the effective spin Hamiltonian of
the TMDL model reads
\begin{equation}
\hat{H}=\frac{\Delta }{2}\sum_{i}\hat{\Sigma}_{i}^{z}-J\sum_{\left\langle
i,j\right\rangle }\left( \hat{\Sigma}_{i}^{x}\hat{\Sigma}_{j}^{x}+\hat{\Sigma%
}_{i}^{y}\hat{\Sigma}_{j}^{y}\right) ,  \label{Hspin}
\end{equation}%
where $\Delta $ is the energy gap between the two states $\left\vert
n\right\rangle $ and $\left\vert n+1\right\rangle $ and acts as a
longitudinal field, and $J=2t(\left\vert \alpha \right\vert ^{2}+\left\vert
\beta \right\vert ^{2})$ is the isotropic exchange interaction. As expected,
we reproduce the continuous $XX$ model even taking the counter-rotating
terms into account.

We now turn to the special case of $N=1$, where only a single Mott lobe
exists. In this case, the mapping procedure of $N\geqslant 2$ can not be
employed directly. However, similar to Ref.~\cite{MSCH12}, the energy gap
between the two lowest energy levels is of higher-order small, compared with
the gap to the next energy level in the ultrastrong coupling regime, and the
numerical calculation verifies that these two lowest levels are still
characterized by two nearest excitation numbers $n$ and $n+1$ (see Appendix
B). Based on these facts, in the ultrastrong coupling regime, we can still
obtain the effective Hamiltonian (\ref{Hspin}) in the subspace spanned by
the two lowest energy levels.\newline

\begin{figure}[tp]
\includegraphics[width=7.5cm]{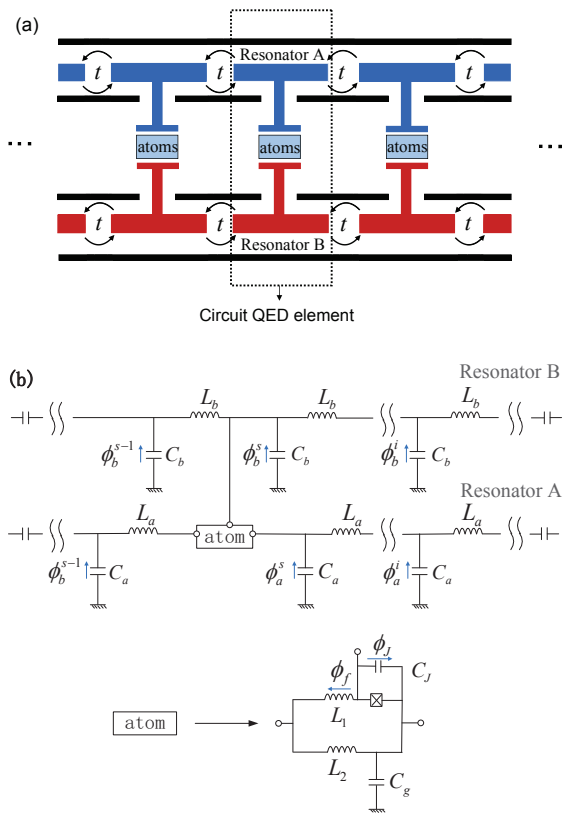}\newline
\caption{(a) Schematic diagram of our proposed two-mode coupled circuit QED
elements (black dashed line), one of which contains a couple of
superconducting stripline resonators and finite Josephson junctions acting
as artificial two-level atoms. The nearest two elements are coupled through
the series capacitance of the resonators with a photon hopping rate $t$. (b)
The effective circuit diagram of each element. The fabricated artificial
atom (black dashed line) is assumed to be placed at a point, which is
labeled by the superscript $s$ of the flux.}
\label{StriplineCircuit}
\end{figure}

\section{Possible experimental implementation}

Having revealed some striking features of the two-mode cavity array, we now
turn to the experimental implementation of the Hamiltonian (\ref{TMDM}).
Motivated by recent experimental achievements of cavity array \cite%
{JR14,CE14,MF16} and multimode cavity \cite{MM11,NMS15,DCM15} in circuit
QED, we propose a scheme, depicted in Fig.~\ref{StriplineCircuit}, to
implement the TMDL model. As shown in Fig.~\ref{StriplineCircuit}(a), the
structure we consider is a series of identical circuit QED elements coupled
through capacities. Each of these elements simulates the single-site
two-mode Dicke model (\ref{TDH}) and the capacitive coupling gives rise to
the photon hopping of different elements. The effective circuit diagram of
each element is shown in Fig.~\ref{StriplineCircuit}(b). A Josephson
junction, acting as an artificial two-level atom, is coupled to two
different superconducting stripline resonators.

We first focus on the circuit QED element labeled in Fig.~\ref%
{StriplineCircuit}(a) with $N=1$. According to the theory of circuit QED, we
can regard the flux $\phi $ and the charge $Q$ as the canonical coordinate
and momentum, respectively. In this sense, the Lagrangian of a circuit QED
element in Fig.~\ref{StriplineCircuit}(b) is written as
\begin{widetext}
\begin{eqnarray}
\mathcal{L} &=&\sum_{i}C_{b}\frac{(\dot{\phi}_{b}^{i})^{2}}{2}+\sum_{i\neq s}%
\left[ C_{a}\frac{(\dot{\phi}_{a}^{i})^{2}}{2}+C_{J}\frac{(\dot{\phi}%
_{J})^{2}}{2}+\tilde{C}_{g}\frac{(\dot{\phi}_{b}^{s}+\dot{\phi}_{J})^{2}}{2}%
\right] -\sum_{i}\frac{(\phi _{b}^{i-1}-\phi _{b}^{i})^{2}}{2L_{b}}
\label{LAG} \\
&&-\sum_{i\neq s}\left[ \frac{(\phi _{a}^{i-1}-\phi _{a}^{i})^{2}}{2L_{a}}-%
\frac{(\phi _{f})^{2}}{2L_{1}}-\frac{(\phi _{f}-\phi _{J})^{2}}{2L_{2}}-%
\frac{(\phi _{a}^{s-1}-\phi _{a}^{s}-\phi _{f}+\phi _{J})^{2}}{2L_{a}}\right]
-E_{J}\cos \left( \frac{\phi _{J}+\phi _{ext}}{\phi _{0}}\right) ,  \notag
\end{eqnarray}%
\end{widetext}where $\tilde{C}_{g}=C_{g}+C_{a}$ and $\phi _{ext}$ is the
external flux of the Josephson junction. Notice that in deriving Eq.~(\ref%
{LAG}), the relation $\dot{\phi}_{a}^{s}=\dot{\phi}_{b}^{s}+\dot{\phi}_{J}$
has been used. Moreover, in terms of the Kirchoff's law at the point, there
exists an extra constraint relation $\phi _{f}=\left(
L_{1}L_{a}+L_{1}L_{2}\right) \phi _{J}/L_{\Sigma }+L_{1}L_{2}\left( \phi
_{a}^{s-1}-\phi _{a}^{s}\right) /L_{\Sigma }$, where $L_{\Sigma
}=L_{2}L_{a}+L_{1}L_{a}+L_{1}L_{2}$.

Using $Q_{k}^{j}=\partial \mathcal{L}/\partial \dot{\phi}_{k}^{j}$, we
obtain the expression of $\dot{\phi}_{k}^{j}$ in terms of $Q_{k}^{j}$, i.e.,
\begin{equation}
\left(
\begin{array}{c}
\dot{\phi}_{J} \\
\dot{\phi}_{b}^{s}%
\end{array}%
\right) =\frac{1}{C_{\Sigma }}\left(
\begin{array}{cc}
C_{b}+\tilde{C}_{g}, & -\tilde{C}_{g} \\
-\tilde{C}_{g}, & C_{J}+\tilde{C}_{g}%
\end{array}%
\right) \left(
\begin{array}{c}
Q_{J} \\
Q_{b}^{s}%
\end{array}%
\right)  \label{PH1}
\end{equation}%
and%
\begin{equation}
\dot{\phi}_{m}^{i\neq s}=\frac{1}{C_{m}}Q_{m}^{i\neq s}\text{ \ (}m=1,2\text{%
)},  \label{PH2}
\end{equation}%
where $C_{\Sigma }=\tilde{C}_{g}C_{b}+\tilde{C}_{g}C_{J}+C_{J}C_{b}$.

By means of Eqs.~(\ref{LAG})-(\ref{PH2}), together with the relation between
the Lagrangian and the Hamiltonian, we expand the Hamiltonian of the circuit
QED element as a sum of three contributions, i.e.,
\begin{equation}
H_{\text{s}}=H_{\text{res}}+H_{\text{at}}+H_{\text{int}}.  \label{HCQ1}
\end{equation}%
In Eq.~(\ref{HCQ1}), the Hamiltonian of the stripline resonator is given by
\begin{eqnarray}
H_{\text{res}}\!&=&\!\sum_{i}\!\left[\!\frac{(Q_{b}^{i})^{2}}{2C_{b}}\!+\!%
\frac{(\phi _{b}^{i}\!-\!\phi _{b}^{i-1})^{2}}{2L_{b}}\frac{(Q_{a}^{i})^{2}}{%
2C_{a}}\!+\!\frac{(\phi _{a}^{i}\!-\!\phi _{a}^{i-1})^{2}}{2L_{a}}\right]
\notag \\
&&+\left( \frac{\tilde{C}_{g}+C_{J}}{2C_{\Sigma }}-\frac{1}{2C_{b}}\right)
(Q_{b}^{s})^{2}-\frac{(Q_{a}^{s})^{2}}{2C_{a}}  \notag \\
&&+\left( \frac{\tilde{L}_{s}}{2L_{\Sigma }^{2}L_{2}L_{a}}-\frac{1}{2L_{a}}%
\right) (\phi _{a}^{s}-\phi _{a}^{s-1})^{2}.  \label{HCQ1a}
\end{eqnarray}%
Since the last three terms in the Hamiltonian (\ref{HCQ1a}) do not involve a
sum over sites, their contributions can be neglected in the continuous
limit, where the number of the sites becomes infinite. Based on this
consideration, we obtain%
\begin{eqnarray}
H_{\text{res}} &=&\sum_{i}\left[ \frac{(Q_{b}^{i})^{2}}{2C_{b}}+\frac{(\phi
_{b}^{i}-\phi _{b}^{i-1})^{2}}{2L_{b}}+\frac{(Q_{a}^{i})^{2}}{2C_{a}}\right.
\label{HCQ1b} \\
&&\left. +\frac{(\phi _{a}^{i}-\phi _{a}^{i-1})^{2}}{2L_{a}}\right] .  \notag
\end{eqnarray}%
The Hamiltonian of the artificial atom reads
\begin{eqnarray}
H_{\text{at}} &=&\frac{\tilde{C}_{g}+C_{b}}{2C_{\Sigma }}(Q_{J})^{2}+\frac{%
\tilde{L}_{J}}{2L_{\Sigma }^{2}L_{2}L_{a}}(\phi _{J})^{2}  \label{Hqu} \\
&&-E_{J}\cos \left( \frac{\phi _{J}+\phi _{ext}}{\phi _{0}}\right) .  \notag
\end{eqnarray}%
The interaction between the artificial atom and the resonator is governed by
the Hamiltonian%
\begin{equation}
H_{\text{int}}=\frac{\tilde{L}_{c}}{2L_{\Sigma }^{2}L_{2}L_{a}}\phi
_{J}(\phi _{a}^{s}-\phi _{a}^{s-1})-\frac{\tilde{C}_{g}}{C_{\Sigma }}%
Q_{J}Q_{b}^{s}.  \label{HCQ1c}
\end{equation}%
In Eqs.~(\ref{HCQ1a})-(\ref{HCQ1c}), $\tilde{L}%
_{J}=L_{1}^{2}L_{2}^{3}+3L_{1}^{2}L_{2}^{2}L_{a}+3L_{1}^{2}L_{2}^{2}L_{a}+3L_{1}^{2}L_{2}L_{a}^{2}+L_{1}^{2}L_{a}^{3}+L_{1}L_{2}^{3}L_{a}+2L_{1}L_{2}^{2}L_{a}^{2}+L_{1}L_{2}L_{a}^{3}-2L_{\Sigma }L_{1}L_{2}^{2}-4L_{\Sigma }L_{1}L_{2}L_{a}-2L_{\Sigma }L_{1}L_{a}^{2}+L_{\Sigma }^{2}L_{2}+L_{\Sigma }^{2}L_{a}
$, $\tilde{L}%
_{s}=L_{1}^{2}L_{2}^{3}+L_{1}^{2}L_{2}^{2}L_{a}+L_{1}L_{2}^{3}L_{a}-2L_{%
\Sigma }L_{1}L_{2}^{2}+L_{\Sigma }^{2}L_{2}$, and $\tilde{L}_{c}=4L_{\Sigma
}L_{1}L_{2}^{2}+4L_{\Sigma
}L_{1}L_{2}L_{a}-2L_{1}^{2}L_{2}^{3}-4L_{1}^{2}L_{2}^{2}L_{a}-2L_{1}^{2}L_{2}L_{a}^{2}-2L_{1}L_{2}^{3}L_{a}-2L_{1}L_{2}^{2}L_{a}^{2}-2L_{\Sigma }^{2}L_{2}
$.

We thus take the continuous limit of the canonical parameters in the
superconducting stripline resonators, i.e., $\phi _{m}^{i}\rightarrow \phi
_{m}(x_{i})$ and $Q_{m}^{i}\rightarrow Q_{m}(x_{i})$, and then promote them
to quantum operators obeying the canonical commutation relation $\left[ \hat{%
\phi}_{m}(x),\hat{Q}_{n}(y)\right] =i\delta (x-y)\delta _{m,n}$. Following
the standard quantization procedure in circuit QED \cite{AB04}, the
quantized canonical parameters are expressed as
\begin{widetext}
\begin{eqnarray}
\hat{\phi}_{m}(x_{i}) &=&\sum_{n_{o}=1}\frac{\sqrt{\omega _{m,n_{o}}L_{m}D}}{%
n_{o}\pi }\cos (\frac{n_{o}\pi x_{i}}{D})(\hat{a}_{m,n_{o}}+\hat{a}%
_{m,n_{o}}^{\dag }) \\
&&+\sum_{n_{e}=2}\frac{\sqrt{\omega _{m,n_{e}}L_{m}D}}{n_{e}\pi }\sin (\frac{%
n_{e}\pi x_{i}}{D})(\hat{a}_{m,n_{e}}+\hat{a}_{m,n_{e}}^{\dag }),  \notag
\end{eqnarray}%
\begin{eqnarray}
\hat{Q}_{m}(x_{i}) &=&-i\sum_{n_{o}=1}\frac{\sqrt{\omega _{m,n_{o}}C_{m}D}}{%
n_{o}\pi }\cos (\frac{n_{o}\pi x_{i}}{D})(\hat{a}_{m,n_{o}}-\hat{a}%
_{m,n_{o}}^{\dag }) \\
&&-i\sum_{n_{e}=2}\frac{\sqrt{\omega _{m,n_{e}}C_{m}D}}{n_{e}\pi }\sin (%
\frac{n_{e}\pi x_{i}}{D})(\hat{a}_{m,n_{e}}-\hat{a}_{m,n_{e}}^{\dag }),
\notag
\end{eqnarray}%
\end{widetext}where $\omega _{m,n}=n\pi /(D\sqrt{L_{m}C_{m}})$ is the
eigenfrequency, $D$ is the length of the resonator, and $n_{o}$ and $n_{e}$
are odd and even integers, respectively.

When the external flux is set to $\phi _{ext}/\phi _{0}=\pi $, the two-level
approximation of the Josephson junction gives that \cite{PC10,AC14}
\begin{equation}
\hat{\phi}_{J}\Leftrightarrow \left\langle \downarrow \right\vert \hat{\phi}%
_{J}\left\vert \uparrow \right\rangle \hat{\sigma}_{x}  \label{PHJ}
\end{equation}%
and
\begin{equation}
\hat{Q}_{J}\Leftrightarrow \frac{\omega _{0}}{4eE_{Q}\phi _{0}}\left\langle
\downarrow \right\vert \hat{\phi}_{J}\left\vert \uparrow \right\rangle \hat{%
\sigma}_{y},  \label{QJ}
\end{equation}%
with $E_{Q}=(\tilde{C}_{g}+C_{b})/(2C_{\Sigma })$, where $\omega _{0}$ is
the resonant frequency of the two-level system, $\left\vert \downarrow
\right\rangle $ and $\left\vert \uparrow \right\rangle $ are the two lowest
macroscopic states of the Hamiltonian $H_{\text{at}}$, and $\hat{\sigma}_{i}$
($i=x,y,z$) is the Pauli spin operator spanned by these two macroscopic
states.

At low temperature, we only keep the mode resonate with the artificial atom
(i.e., $n=1$) and neglect other non-resonate terms. Under this single-mode
approximation of the resonator and the two-level approximation of the
artificial atom, the Hamiltonian of the considered circuit QED element is
finally expressed as%
\begin{eqnarray}
\hat{H}_{s} &=&\omega _{1}\hat{a}_{1}^{\dag }\hat{a}_{1}+\omega _{2}\hat{a}%
_{2}^{\dag }\hat{a}_{2}+\frac{1}{2}\omega _{0}\hat{\sigma}_{z}  \label{HS} \\
&&+g_{1}(\hat{a}_{1}+\hat{a}_{1}^{\dag })\hat{\sigma}_{x}+ig_{2}(\hat{a}_{2}-%
\hat{a}_{2}^{\dag })\hat{\sigma}_{y},  \notag
\end{eqnarray}%
where
\begin{equation}
g_{1}=-\frac{\tilde{L}_{c}\sqrt{\omega _{1}/L_{a}D}\sin (\pi
x_{s}/D)\left\langle \downarrow \right\vert \hat{\phi}_{J}\left\vert
\uparrow \right\rangle }{2L_{\Sigma }^{2}L_{2}},  \label{g1}
\end{equation}%
\begin{equation}
g_{2}=\frac{\tilde{C}_{g}\omega _{0}\sqrt{\omega _{2}C_{b}D}\cos (\pi
x_{s}/D)\left\langle \downarrow \right\vert \hat{\phi}_{J}\left\vert
\uparrow \right\rangle }{4\pi eE_{Q}\phi _{0}C_{\Sigma }},  \label{g2}
\end{equation}%
\begin{equation}
\omega _{1}=\frac{\pi }{D\sqrt{L_{a}C_{a}}},  \label{O1}
\end{equation}%
\begin{equation}
\omega _{2}=\frac{\pi }{D\sqrt{L_{b}C_{b}}}.  \label{O2}
\end{equation}%
The tunability of the inductance and the capacitance of the two
superconducting stipline resonators allows us to set $\omega _{1}=\omega
_{2}=\omega $ and $g_{1}=g_{2}=g_{0}$, under which the Hamiltonian (\ref{HS}%
) reduces to the single-site two-mode Rabi model. Using the same procedure,
the Hamiltonian (\ref{HS}) can be extended straightforwardly to the case
with several two-level artificial atoms, i.e., the single-site two-mode
Dicke Hamiltonian (\ref{TDH}). When a series of such circuit QED elements
are coupled capacitively with the hopping rate $t$ [see Fig.~\ref%
{StriplineCircuit}(a)], the TMDL Hamiltonian (\ref{TMDM}) can be achieved.

We emphasize that the improvement of current experimental techniques in the
ultrastrong-coupling circuit QED \cite{TN10,PF10,PJ16,FY16} makes our
proposal a promising candidate to exhibit relevant physics of the TMDL model.%
\newline

\section{Discussions}

\begin{figure}[tp]
\includegraphics[width=7.9cm]{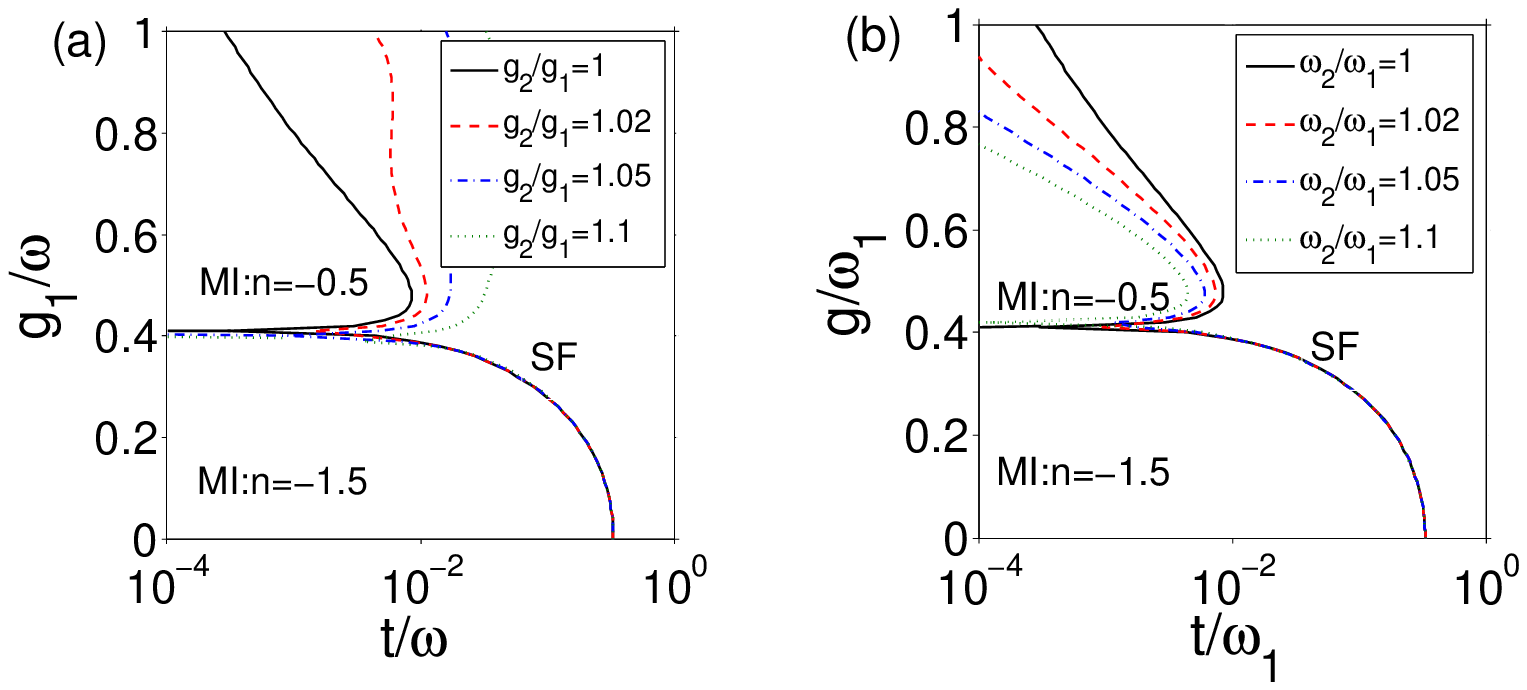}\newline
\caption{(a) Phase boundaries for $g_{2}/g_{1}=1$ (black solid curve), $%
g_{2}/g_{1}=1.02$ (red dashed curve), $g_{2}/g_{1}=1.05$ (blue dotted-dashed
curve), and $g_{2}/g_{1}=1.1$ (green dotted curve), when $\protect\omega %
_{1}/\protect\omega =\protect\omega _{2}/\protect\omega =\protect\omega _{0}/%
\protect\omega =1$. (b) Phase boundaries for $\protect\omega _{2}/\protect%
\omega _{1}=1$ (black solid curve), $\protect\omega _{2}/\protect\omega %
_{1}=1.02$ (red dashed curve), $\protect\omega _{2}/\protect\omega _{1}=1.05$
(blue dotted-dashed curve), and $\protect\omega _{2}/\protect\omega _{1}=1.1$
(green dotted curve), when $g_{1}/g=g_{2}/g=\protect\omega _{0}/\protect%
\omega _{1}=1$. In these figures, we set $N=3$.}
\label{DeviPhase}
\end{figure}

Up to now, our discussions are restricted to the case of the degenerate
photon modes ($\omega _{1}=\omega _{2}=\omega $) and the equal atom-photon
coupling strengths ($g_{1}=g_{2}=g$). If these conditions are not fulfilled,
there would not be a strict conservation law of $\hat{N}_{e}$, and an
instructive question is whether the Mott-lobe structure still exists in such
a case or not. To briefly show the influence of a slight deviation of these
two equalities, $\omega _{1}=\omega _{2}=\omega $ and $g_{1}=g_{2}=g$, we
plot the phase diagrams in the $t-g$ plane for different $\omega _{2}/\omega
_{1}$ [Fig.~\ref{DeviPhase}(a)] or $g_{2}/g_{1}$ [Fig.~\ref{DeviPhase}(b)],
when $N=3$. It can be seen clearly from these figures that a slight
deviation of the ideal condition does not break the Mott-lobe structure but
merely shift the phase boundary.\newline

\section{Conclusion}

In summary, we have constructed a new type of cavity array system, which is
governed by the TMDL model. This model incorporates all of the
counter-rotating terms of the atom-photon coupling and therefore works well
in the ultrastrong coupling regime achieved in recent experiments. Unlike
the standard Dicke-lattice model, the TMDL has a global conserved excitation
and a continuous $U(1)$ symmetry. This distinct change of symmetry strongly
impacts the nature of photon localization/delocalization behavior.
Specifically, the atom-photon interaction features Mott-lobe structures of
photons and a second-order SF-MI phase transition, which share similarities
with the Jaynes-Cummings-lattice and Bose-Hubbard models. However, the
Mott-lobe structures predicted here depend crucially on the atom number of
each site, reflecting its particularity among lattice models. We have also
shown that the TMDL model can be mapped into a continuous $XX$ spin model
under proper parameter conditions. Finally, we have proposed an
experimentally-feasible scheme to realize the TMDL model in a two-mode
superconducting stripline cavity array.\newline

\section{Acknowledgements}

This work is supported partly by the NSFC under Grants No.~11422433,
No.~11674200, No.~11604392, No.~11434007, and No.~61378049; the FANEDD under
Grant No.~201316; SFSSSP; OYTPSP; and SSCC.\newline

\vbox{\vskip1cm} \appendix
\setcounter{figure}{0} \renewcommand{\thefigure}{B\arabic{figure}}

\section{Mapping $\hat{a}_{1,j}$ and $\hat{a}_{2,j}$ to\ the spin operators}

We first notice that the commutation relations between the photon
annihilation operator and the excitation number operator satisfy $\left[
\hat{a}_{1,j},\hat{N}_{e,j}\right] =\hat{a}_{2,j}$ and $\left[ \hat{a}_{2,j},%
\hat{N}_{e,j}\right] =\hat{a}_{1,j}$. Taking these two equations into
account, the matrix elements of $\hat{a}_{1,j}+\hat{a}_{2,j}$ and $\hat{a}%
_{1,j}-\hat{a}_{2,j}$ in the basis of the excitation eigenstates $\left\vert
n\right\rangle $ and $\left\vert m\right\rangle $ are expressed respectively
as%
\begin{eqnarray}
\left\langle n\right\vert \hat{a}_{1,j}\!\!+\!\!\hat{a}_{2,j}\left\vert
m\right\rangle &\!\!\!=\!\!\!&\left\langle n\right\vert \left[ \hat{a}_{1,j}+%
\hat{a}_{2,j},\hat{N}_{e,j}\right] \left\vert m\right\rangle \\
&=&(m-n)\left\langle n\right\vert \hat{a}_{1,j}+\hat{a}_{2,j}\left\vert
m\right\rangle ,  \notag \\
\left\langle n\right\vert a_{1,j}\!\!-\!\!a_{2,j}\left\vert m\right\rangle
&=&-\left\langle n\right\vert \left[ \hat{a}_{1,j}\!-\!\hat{a}_{2,j},N_{e,j}%
\right] \left\vert m\right\rangle \\
&=&(n-m)\left\langle n\right\vert \hat{a}_{1,j}-\hat{a}_{2,j}\left\vert
m\right\rangle .  \notag
\end{eqnarray}%
To obtain a nonzero value of $\left\langle n\right\vert \hat{a}_{1,j}+\hat{a}%
_{2,j}\left\vert m\right\rangle $ ($\left\langle n\right\vert \hat{a}_{1,j}-%
\hat{a}_{2,j}\left\vert m\right\rangle $), we should have $m=n+1$ ($m=n-1$),
and in the reduced Hilbert space $\left\{ \left\vert n\right\rangle
,\left\vert n+1\right\rangle \right\} $, the operators $\hat{a}_{1,j}+\hat{a}%
_{2,j}$ and $\hat{a}_{1,j}-\hat{a}_{2,j}$ thus read%
\begin{eqnarray}
\hat{a}_{1,j}+\hat{a}_{2,j} &\rightarrow &2\alpha \left(
\begin{array}{cc}
0 & 0 \\
1 & 0%
\end{array}%
\right) \Longleftrightarrow 2\alpha \hat{\Sigma}_{j}^{-},\text{ } \\
\hat{a}_{1,j}-\hat{a}_{2,j} &\rightarrow &2\beta \left(
\begin{array}{cc}
0 & 1 \\
0 & 0%
\end{array}%
\right) \Longleftrightarrow 2\beta \hat{\Sigma}_{j}^{+},
\end{eqnarray}%
from which we can straightforwardly obtain $\hat{a}_{1,j}\rightarrow $ $%
\alpha \hat{\Sigma}_{j}^{-}+\beta \hat{\Sigma}_{j}^{+}$ and $%
a_{2,j}\rightarrow $ $\alpha \hat{\Sigma}_{j}^{-}-\beta \hat{\Sigma}_{j}^{+}$%
, i.e., Eq.~(\ref{OP}) of the main text. The coefficients $\alpha $ and $%
\beta $ can be determined numerically.

\section{Numerical demonstration of the two state subspace $\left\{
\left\vert n\right\rangle ,\left\vert n+1\right\rangle \right\} $ in the
ultrastrong coupling regime for $N=1$}

Figure \ref{TOCD_Energylevel} shows the low-lying spectrum of the
Hamiltonian (\ref{TDH}) with $N=1$, from which we can see clearly that the
two lowest energy levels become quasi-degenerate in the ultrastrong coupling
regime. Moreover, as shown in the inset of this figure, both of these two
levels support the well-defined excitation numbers, whose difference remains
one. This guarantees the validity of the truncation of the Hilbert space to
an effective two state subspace $\left\{ \left\vert n\right\rangle
,\left\vert n+1\right\rangle \right\} $ for a large $g/\omega $.

\begin{figure}[tp]
\includegraphics[width=7cm]{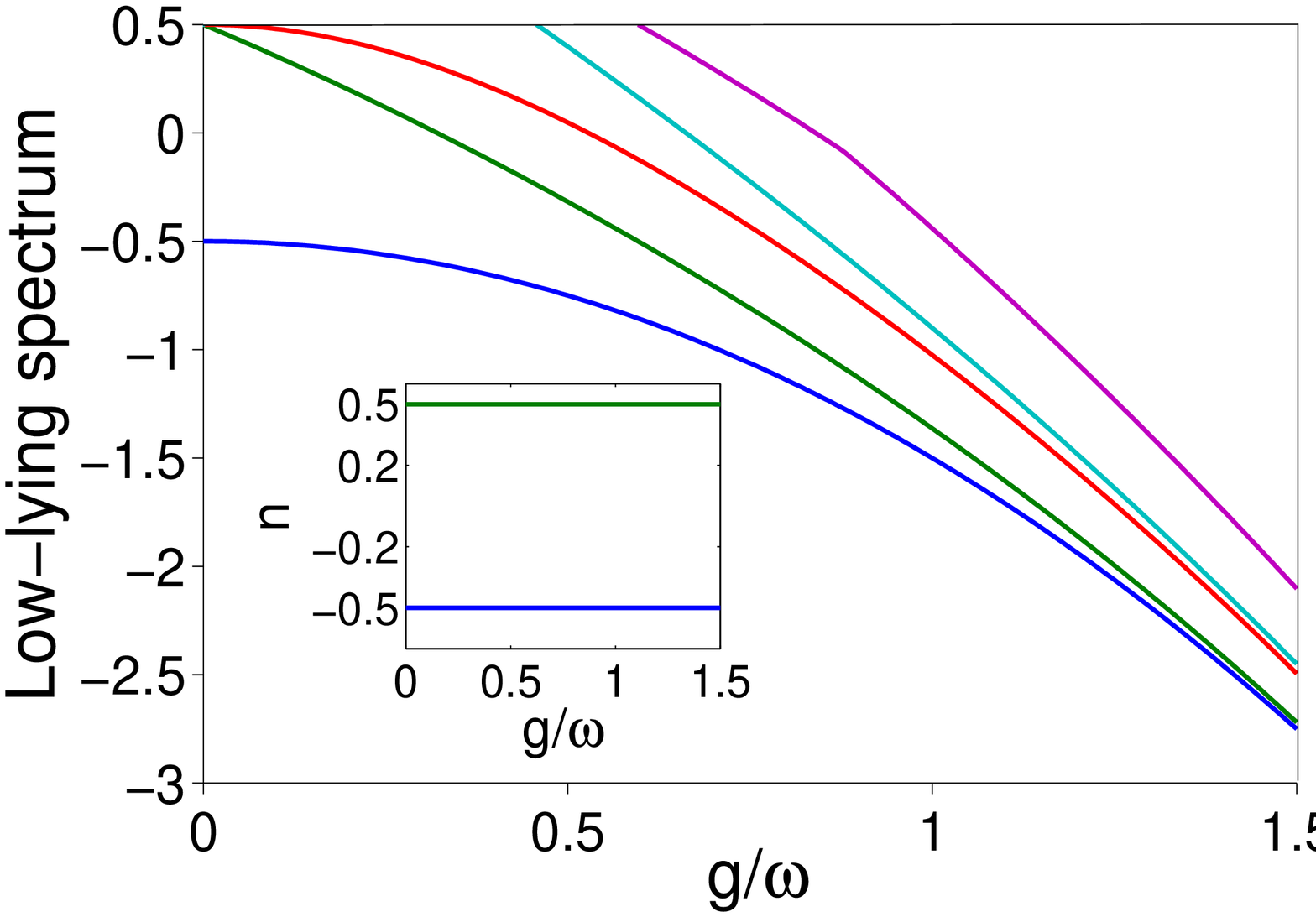}\newline
\caption{The low-lying spectrum of the Hamiltonian (\protect\ref{TDH}) as a
function of $g/\protect\omega $. Inset: the excitation numbers $n$ of the
two lowest levels. In these figures, we set $\protect\omega _{0}/\protect%
\omega =1$ and $N=1$}
\label{TOCDEnergylevel}
\end{figure}


\begin{thebibliography}{99}
\bibitem{TEN14} T. E. Northup and R. Blatt, Quantum information transfer
using photons, Nat. Photon. \textbf{8}, 356 (2014).

\bibitem{IC13} I. Carusotto and C. Ciuti, Quantum fluids of light, Rev. Mod.
Phys. \textbf{85}, 299 (2013).

\bibitem{MJHF08} M. J. Hartmann, F. G. S. L. Brand\~{a}o, and M. B. Plenio,
Quantum many-body phenomena in coupled cavity arrays, Laser Photon. Rev.
\textbf{2}, 527 (2008).

\bibitem{AAH12} A. A. Houck, H. E. T\"{u}reci, and J. Koch, On-chip quantum
simulation with superconducting circuits, Nat. Phys. \textbf{8}, 292 (2012).

\bibitem{SS13} S. Schmidt and J. Koch, Circuit QED lattices: Towards quantum
simulation with superconducting circuits, Ann. Phys. \textbf{525}, 395 (2013)

\bibitem{IMG14} I. M. Georgescu, S. Ashhab, and F. Nori, Quantum simulation,
Rev. Mod. Phys. \textbf{86}, 153 (2014).

\bibitem{CN17} C. Noh and D. G Angelakis, Quantum simulations and many-body
physics with light, Rep. Prog. Phys. \textbf{80}, 016401 (2017).

\bibitem{ADG06} A. D. Greentree, C. Tahan, J. H. Cole, and L. C.
L.Hollenberg, Quantum phase transiton of light, Nat. Phys. \textbf{2}, 856
(2006).

\bibitem{MFA06} M. J. Hartmann, F. G. S. L. Brand\~{a}o, and M. B. Plenio,
Strongly interacting polaritons in coupled arrays of cavities, Nat. Phys.
\textbf{2}, 849 (2006).

\bibitem{MJH07} M. J. Hartmann, F. G. S. L. Brand\~{a}o, and M. B. Plenio,
Effective Spin Systems in Coupled Microcavities, Phys. Rev. Lett. \textbf{99}%
, 160501 (2007).

\bibitem{DGA07} D. G. Angelakis, M. F. Santos, and S. Bose,
Photon-blockade-induced Mott transitions and XY spin models in coupled
cavity arrays, Phys. Rev. A \textbf{76}, 031805(R) (2007).

\bibitem{MIM08} M. I. Makin, J. H. Cole, C, Tahan, L. C. L. Hollenberg, and
A. D. Greentree, Quantum phase transitions in photonic cavities with
two-level systems, Phys. Rev. A \textbf{77}, 053819 (2008).

\bibitem{SS09} S. Schmidt and G. Blatter, Strong Coupling Theory for the
Jaynes-Cummings-Hubbard Model, Phys. Rev. Lett. \textbf{103}, 086403 (2009)

\bibitem{JK09} J. Koch and K. L. Hur, Superfluid-Mott-insulator transition
of light in the Jaynes-Cummings lattice, Phys. Rev. A \textbf{80}, 023811
(2009).

\bibitem{KT13} K. Toyoda, Y. Matsuno, A. Noguchi, S. Haze, and S. Urabe,
Experimental Realization of a Quantum Phase Transition of Polaritonic
Excitations, Phys. Rev. Lett. \textbf{111}, 160501 (2013).

\bibitem{GK13} G. Kulaitis, F. Kr\"{u}ger, F. Nissen, and J. Keeling,
Disordered driven coupled cavity arrays: Nonequilibrium stochastic
mean-field theory, Phys. Rev. A \textbf{87}, 013840 (2013).

\bibitem{KKM13} K. Kamide, M. Yamaguchi, T. Kimura, and T. Ogawa,
First-order superfluid-Mott-insulator transition for quantum-optical
switching in cavity-QED arrays with two cavity modes, Phys. Rev. A \textbf{87%
}, 053842 (2013).

\bibitem{SF14} S. Felicetti, G. Romero, D. Rossini, R. Fazio, and E. Solano,
Photon transfer in ultrastrongly coupled three-cavity arrays, Phys. Rev. A
\textbf{89}, 013853 (2014).

\bibitem{BB14} B. Bujnowski, J. K. Corso, A. L. C. Hayward, J. H. Cole, and
A. M. Martin, Supersolid phases of light in extended Jaynes-Cummings-Hubbard
systems, Phys. Rev. A \textbf{90}, 043801 (2014).

\bibitem{MB15} M. Biondi, E. P. L. van Nieuwenburg, G. Blatter, S. D. Huber,
and S. Schmidt, Incompressible Polaritons in a Flat Band, Phys. Rev. Lett.
\textbf{115}, 143601 (2015).

\bibitem{ZY15} Y. Zhang, J. Fan, J. -Q. Liang, J. Ma, G. Chen, S. Jia, and
F. Nori, Photon devil's staircase: photon long-range repulsive interaction
in lattices of coupled resonators with Rydberg atoms, Sci. Rep. \textbf{5},
11510 (2015).

\bibitem{KS15} K. Seo and L. Tian, Quantum phase transition in a
multiconnected superconducting Jaynes-Cummings lattice, Phys. Rev. B \textbf{%
91}, 195439 (2015).

\bibitem{MS16} M. Schir\'{o}, C. Joshi, M. Bordyuh, R. Fazio, J. Keeling,
and H. E. T\"{u}reci, Exotic Attractors of the Nonequilibrium Rabi-Hubbard
Model, Phys. Rev. Lett. \textbf{116}, 143603 (2016).

\bibitem{HMJ16} M.-J. Hwang and M. B. Plenio, Quantum Phase Transition in
the Finite Jaynes-Cummings Lattice Systems, Phys. Rev. Lett. \textbf{117},
123602 (2016).

\bibitem{AM16} A. Maggitti, M. Radonji\'{c}, and B. M. Jelenkovi\'{c},
Dark-polariton bound pairs in the modified Jaynes-Cummings-Hubbard model,
Phys. Rev. A \textbf{93}, 013835 (2016).

\bibitem{ALCH16} A. L. C. Hayward and A. M. Martin, Superfluid-Mott
transitions and vortices in the Jaynes-Cummings-Hubbard lattices with
time-reversal-symmetry breaking, Phys. Rev. A \textbf{93}, 023828 (2016).

\bibitem{MPA89} M. P. A. Fisher, P. B. Weichman, G. Grinstein, and D. S.
Fisher, Boson localization and the superfluid-insulator transition, Phys.
Rev. B \textbf{40}, 546 (1989).

\bibitem{IB08} I. Bloch, J. Dalibard, and W. Zwerger, Many-body physics with
ultracold gases, Rev. Mod. Phys. \textbf{80}, 885 (2008).

\bibitem{TN10} T. Niemczyk, F. Deppe, H. Huebl, E. P. Menzel, F. Hocke, M.
J. Schwarz, J. J. Garc\'{\i}a-Ripoll, D. Zueco, T. H\"{u}mmer, E. Solano, A.
Marx, and R. Gross, Circuit quantum electrodynamics in the
ultrastrong-coupling regime, Nat. Phys. \textbf{6}, 772 (2010).

\bibitem{PF10} P. Forn-D\'{\i}az, J. Lisenfeld, D. Marcos, J. J. Garc\'{\i}%
a-Ripoll, E. Solano, C. J. P. M. Harmans, and J. E. Mooij, Observation of
the Bloch-Siegert Shift in a Qubit-Oscillator System in the Ultrastrong
Coupling Regime, Phys. Rev. Lett. \textbf{105}, 237001 (2010).

\bibitem{PJ16} P. Forn-D\'{\i}az, J. J. Garc\'{\i}a-Ripoll, B. Peropadre,
J.-L. Orgiazzi, M. A. Yurtalan, R. Belyansky, C. M. Wilson, and A. Lupascu,
Ultrastrong coupling of a single artificial atom to an electromagnetic
continuum in the nonperturbative regime, Nat. Phys. doi:10.1038/nphys3905
(2016).

\bibitem{FY16} F. Yoshihara, T. Fuse, S. Ashhab, K. Kakuyanagi, S. Saito,
and K. Semba, Superconducting qubit--oscillator circuit beyond the
ultrastrong-coupling regime, Nat. Phys. doi:10.1038/nphys3906 (2016).

\bibitem{HZ11} H. Zheng and Y. Takada, Importance of counter-rotating
coupling in the superfluid-to-Mott-insulator quantum phase transition of
light in the Jaynes-Cummings lattice, Phys. Rev. A \textbf{84}, 043819
(2011).

\bibitem{MSCH12} M. Schir\'{o}, M. Bordyuh, B. \"{O}ztop, and H. E. T\"{u}%
reci, Phase Transition of Light in Cavity QED Lattices, Phys. Rev. Lett.
\textbf{109}, 053601 (2012).

\bibitem{MSCH13} M. Schir\'{o}, M. Bordyuh, B. \"{O}ztop, and H. E. T\"{u}%
reci, Quantum phase transition of light in the Rabi-Hubbard model, J. Phys.
B \textbf{46}, 224021 (2013).

\bibitem{TF16} T. Flottat, F. H\'{e}bert, V. G. Rousseau, and G. G.
Batrouni, Quantum Monte Carlo study of the Rabi-Hubbard model, Eur. Phys. J.
D \textbf{70}, 213 (2016).

\bibitem{BS13} B. Kumar and S. Jalal, Quantum Ising dynamics and
Majorana-like edge modes in the Rabi lattice model, Phys. Rev. A \textbf{88}%
, 011802(R) (2013).

\bibitem{JR14} J. Raftery, D. Sadri, S. Schmidt, H. E. T\"{u}reci, and A. A.
Houck, Observation of a Dissipation-Induced Classical to Quantum Transition,
Phys. Rev. X \textbf{4}, 031043 (2014).

\bibitem{CE14} C. Eichler, Y. Salathe, J. Mlynek, S. Schmidt, and A.
Wallraff, Quantum-Limited Amplification and Entanglement in Coupled
Nonlinear Resonators, Phys. Rev. Lett. \textbf{113}, 110502 (2014).

\bibitem{MF16} M. Fitzpatrick, N. Sundaresan, A. C. Y. Li, J. Koch, and A.
A. Houck, Observation of a dissipative phase transition in a one-dimensional
circuit QED lattice, arXiv 1607.06895v1 (2016).

\bibitem{MM11} M. Mariantoni, H. Wang, R. C. Bialczak, M. Lenander, E.
Lucero, M. Neeley, A. D. O'Connell, D. Sank, M. Weides, J. Wenner, T.
Yamamoto, Y. Yin, J. Zhao, J. M. Martinis, and A. N. Cleland, Photon shell
game in three-resonator circuit quantum electrodynamics, Nat. Phys. \textbf{7%
}, 287 (2011).

\bibitem{NMS15} N. M. Sundaresan, Y. Liu, Darius S., L\'{a}szl\'{o}, J. Sz%
\H{o}cs, D. L. Underwood, M. Malekakhlagh, H. E. T\"{u}reci, and A. A.
Houck, Beyond Strong Coupling in a Multimode Cavity, Phys. Rev. X \textbf{5}%
, 021035 (2015).

\bibitem{DCM15} D. C. McKay, R. Naik, P. Reinhold, L. S. Bishop, and D. I.
Schuster, High-Contrast Qubit Interactions Using Multimode Cavity QED, Phys.
Rev. Lett. \textbf{114}, 080501 (2015).

\bibitem{LJZ14} L. J. Zou, D. Marcos, S. Diehl, S. Putz, J. Schmiedmayer, J.
Majer, and P. Rabl, Implementation of the Dicke Lattice Model in Hybrid
Quantum System Arrays, Phys. Rev. Lett. \textbf{113}, 023603 (2014).

\bibitem{JF14} J. Fan, Z. Yang, Y. Zhang, J. Ma, G. Chen, and S. Jia, Hidden
continuous symmetry and Nambu-Goldstone mode in a two-mode Dicke model,
Phys. Rev. A \textbf{89}, 023812 (2014).

\bibitem{CT03} C. Emary and T. Brandes, Chaos and the quantum phase
transition in the Dicke model, Phys. Rev. E \textbf{67}, 066203 (2003).


\bibitem{SCR08} S.-C. Lei and R.-K. Lee, Quantum phase transition of light
in the Dicke-Bose-Hubbard model, Phys. Rev. A \textbf{77}, 033827 (2008).

\bibitem{JPS06} J. P. Sethna, \textit{Statistical Mechanics: Entropy, Order
Parameters, and Complexity} (Oxford University Press, New York, 2006).

\bibitem{KH87} K. Huang, \textit{Statistical Mechanics} (Wiley, New York,
1987).

\bibitem{DR07} D. Rossini and R. Fazio, Mott-Insulating and Glassy Phases of
Polaritons in 1D Arrays of Coupled Cavities, Phys. Rev. Lett. \textbf{99},
186401 (2007).

\bibitem{MK10} M. Knap, E. Arrigoni, and W. von der Linden, Quantum phase
transition and excitations of the Tavis-Cummings lattice model, Phys. Rev. B
\textbf{82}, 045126 (2010).

\bibitem{AB04} A. Blais, R.-S. Huang, A. Wallraff, S. M. Girvin, and R. J.
Schoelkopf, Cavity quantum electrodynamics for superconducting electrical
circuits: An architecture for quantum computation, Phys. Rev. A \textbf{69},
062320 (2004).

\bibitem{PC10} P. Nataf and C. Ciuti, Vacuum Degeneracy of a Circuit QED
System in the Ultrastrong Coupling Regime, Phys. Rev. Lett. \textbf{104},
023601 (2010).

\bibitem{AC14} A. Baksic and C. Ciuti, Controlling Discrete and Continuous
Symmetries in \textquotedblleft Superradiant\textquotedblright\ Phase
Transitions with Circuit QED Systems, Phys. Rev. Lett. \textbf{112}, 173601
(2014).
\end{thebibliography}
\end{document}